\documentclass[11pt]{cernrep}
\usepackage{graphicx}
\usepackage{here}

\begin{document}

\title{Confronting classical and Bayesian confidence limits to
examples\footnote{Contribution to the Workshop on Confidence Limits held at
CERN, January 2000.}}
\author{G\"{u}nter Zech\thanks{E-mail: zech@physik.uni-siegen.de}\\Universit\"{a}t Siegen, D-57068 Siegen}
\maketitle
\begin{abstract}
Classical confidence limits are compared to Bayesian error bounds by studying
relevant examples. The performance of the two methods is investigated relative
to the properties coherence, precision, bias, universality, simplicity. A
proposal to define error limits in various cases is derived from the
comparison. It is based on the likelihood function only and follows in most
cases the general practice in high energy physics. Classical methods are
discarded because they violate the likelihood principle, they can produce
physically inconsistent results, suffer from a lack of precision and
generality. Also the extreme Bayesian approach with arbitrary choice of the
prior probability density or priors deduced from scaling laws is rejected.
\end{abstract}

\section{Purpose, criteria, definitions}

The progress of experimental sciences to a large extent is due to their
practice to assign uncertainties to results. The information contained in a
measurement, or a parameter deduced from it, is incompletely documented and
more or less useless unless some kind of error is attributed to the data. The
precision of measurements has to be known i) to combine data from different
experiments, ii) to deduce secondary parameters from it and iii) to test
predictions of theories. Different statistical methods have to be judged on
their ability to fulfill these tasks.

Narsky \cite{nars99} who compares several different approaches to the
estimation of upper Poisson limits, states: ``There is no such thing as the
best procedure for upper limit estimation. An experimentalist is free to
choose any procedure she/he likes, based on her/his belief and experience. The
only requirement is that the chosen procedure must have a strict mathematical
foundation.'' This opinion is typical for many papers on confidence limits.
However, ``the real test of the pudding is in its eating'' and not in
contemplating the beauty of the cooking recipe. We should not forget that what
we measure has practical implications.

In this paper, the emphasis is put on performance and not on the mathematical
and statistical foundation. The intention is to confront the procedures with
the problems to be solved in physics. Simple transparent examples are
selected. Important properties are among others consistency, precision,
universality, simplicity and objectivity.

Consistency is indispensable in any case. A. W. F. Edwards writes
\cite{edwa92}: ``Relative support (of a hypothesis or a parameter) must be
consistent in different applications, so that we are content to react equally
to equal values, and it must not be affected by information judged intuitively
to be irrelevant.''

Part of the content of this article has been presented in a comment
\cite{zech98} to the unified approach \cite{feld98}.

\subsection{Classical confidence limits}

Classical confidence limits (CCL) are based on tail probabilities. The
defining property is \emph{coverage}: If a large number of experiments perform
measurements of a parameter with confidence level $\alpha$, the fraction
$\alpha$ of the limits will contain the true value of the parameter inside the
confidence limits.%

\begin{figure}
\begin{center}
\includegraphics[
width=4.9787in
]%
{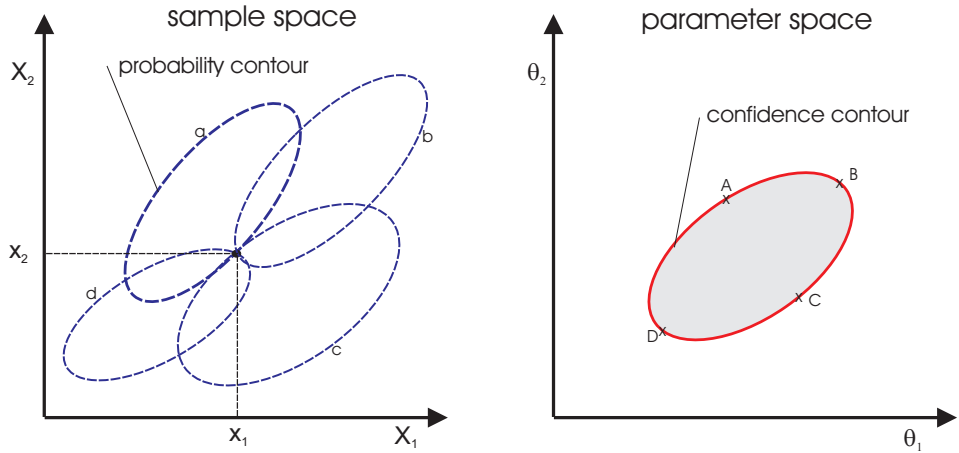}%
\caption{{\small Two parameter classical confidence limit for a measurement
}$x_{1,}x_{2}${\small . The dashed contours labeled with small letters in the
sample space correspond to probability contours of the parameter pairs labeled
with capital letters in the parameter space. }}%
\end{center}
\end{figure}

We illustrate the concept of CCL for a measurement (statistic) consisting of a
two-dimensional observation ($x_{1},x_{2}$) and a two dimensional parameter
space (see Fig. 1). In a first step we associate to each point $\theta
_{1},\theta_{2}$ in the parameter space a closed \emph{probability contour} in
the sample space containing a measurement with probability $\alpha$. For
example, the probability contour labeled $a$ in the sample space corresponds
to the parameter values of point $A$ in the parameter space. The curve
(\emph{confidence contour}) connecting all points in the parameter space with
probability contours in the sample space passing through the actual
measurement $x_{1},x_{2}$ encloses the \emph{confidence region} of confidence
level $\alpha$.

Figure 1 demonstrates some of the requirements necessary for the
construction of an exact confidence region: 1. The sample space must be
continuous. (Discrete distributions and thus all digital measurements and in
principle also Poisson processes are excluded.) 2. The probability contours
should enclose a simply connected region. 3. The parameter space has to be continuous.

The restriction (1) usually is overcome by relaxing the requirement of exact
coverage and by requiring minimum overcoverage. This is not an elegant solution.

There is considerable freedom in the choice of the probability contours but to
insure coverage they have to be defined independently of the result of the
experiment. Usually, contours are locations of constant probability density.
In one dimension also central intervals and intervals leading to minimum sized
confidence intervals are popular. Clearly, there is a lack of standardization.
The unified approach \cite{feld98} defines the probability regions through the
likelihood ratio.

\subsection{Likelihood limits and Bayesian conventions}

Likelihood intervals enclose a region where the likelihood function decreases
by a fixed ratio, equal to $\sqrt{e}$ for one standard deviation and $e^{2}$
for two standard deviations etc..%

\begin{figure}
\begin{center}
\includegraphics[width=13.5cm
]%
{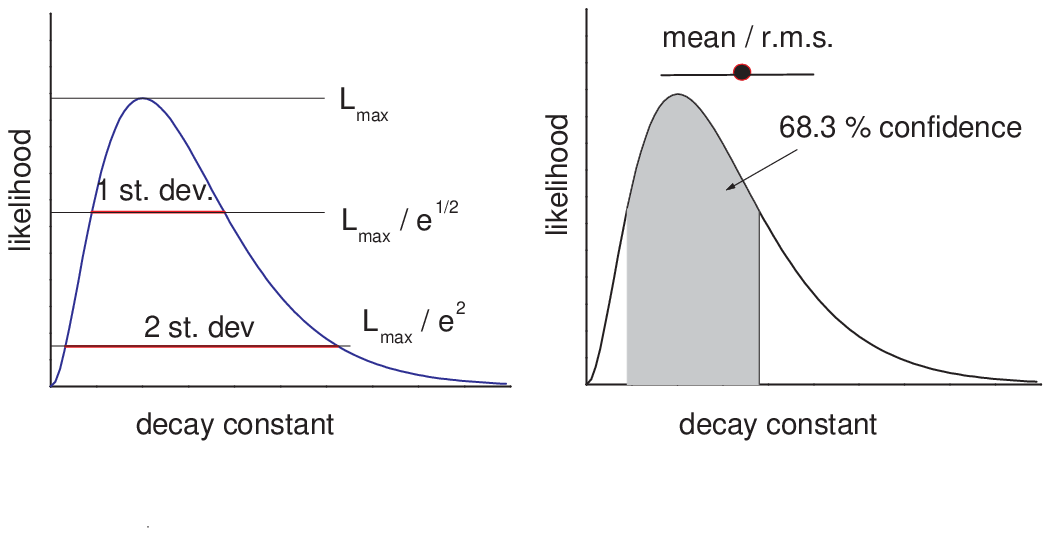}%
\caption{{\small Likelihood limits (left) and Bayesian limits (right).}}%
\end{center}
\end{figure}

Bayesians integrate the normalized likelihood function and form either
probability regions or moments to define the limits. I will discuss only
uniform prior densities. This does not restrict the freedom of the scientist
because there is the equivalent possibility to choose the parameter. For
example an analysis using the mean life parameter with the prior $1/\tau^{2}$
is equivalent to an analysis of the decay constant $\gamma$ with uniform prior.

\subsection{The likelihood principle}

Assume we have two hypothesis characterized by the parameters $\theta_{1}$ and
$\theta_{2}$. For a measurement $x_{1}$ the relative support to the two
hypothesis is given by the likelihood ratio
\[
R(x_{1}|\theta_{1},\theta_{2})=\frac{P(x_{1}|\theta_{1})}{P(x_{1}|\theta_{2})}%
\]
Another measurement $x_{2}$ is equivalent to $x_{1}$ if the likelihood ratios
are the same:
\[
\frac{P(x_{1}|\theta_{1})}{P(x_{1}|\theta_{2})}=\frac{P(x_{2}|\theta_{1}%
)}{P(x_{2}|\theta_{2})}%
\]

When we have more than two hypothesis we require that equivalent date provide
the same likelihood ratio for all combinations of parameters. Consequently,
for a pdf depending on a continuous parameter $\theta,$ we have to require
that the likelihood functions for the two measurements are proportional to
each other. These considerations correspond to the Likelihood Principle (LP):
The likelihood function contains the full information relative to the
parameter. Inference should be based on the likelihood function only. The LP
is due to Fisher, Birnbaum and others. Proofs and discussions can be found in
Refs. \cite{basu88, berg84, edwa92}.

Methods that provide different results for measurement that have proportional
likelihood functions are inconsistent.

\section{Examples}

\subsection{Example 1a: Gaussian with physical boundary}

\bigskip%
\begin{figure}
[ptb]
\begin{center}
\includegraphics[width=14.5cm
]%
{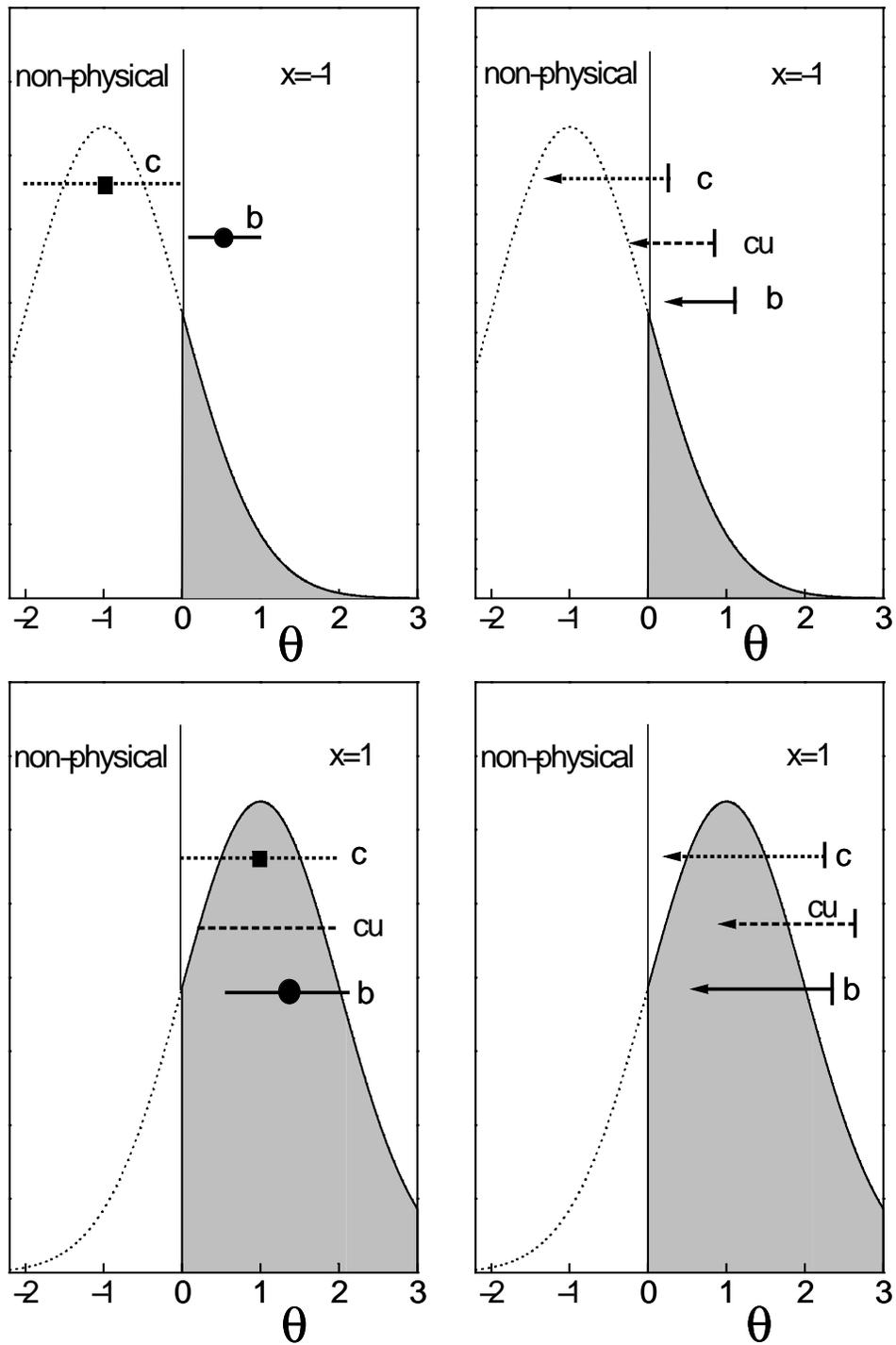}%
\caption{{\small Gaussian errors near physical boundary (c: classical, cu:
unified, l: likelihood, b: Bayesian). Left: 68.3\% errors, right:90\% upper
limits.}}%
\end{center}
\end{figure}

A physical quantity like the mass of a particle with a resolution following
normal distributions is constrained to positive values. Fig. 3 shows typical
central confidence bounds which extend into the unphysical region. In extreme
cases a measurement may produce a 90 \% confidence interval which does not
cover positive values at all. The unified approach and the Bayesian method
avoid unphysical confidence limits.%

\begin{figure}
\begin{center}
\includegraphics[width=8.5cm
]%
{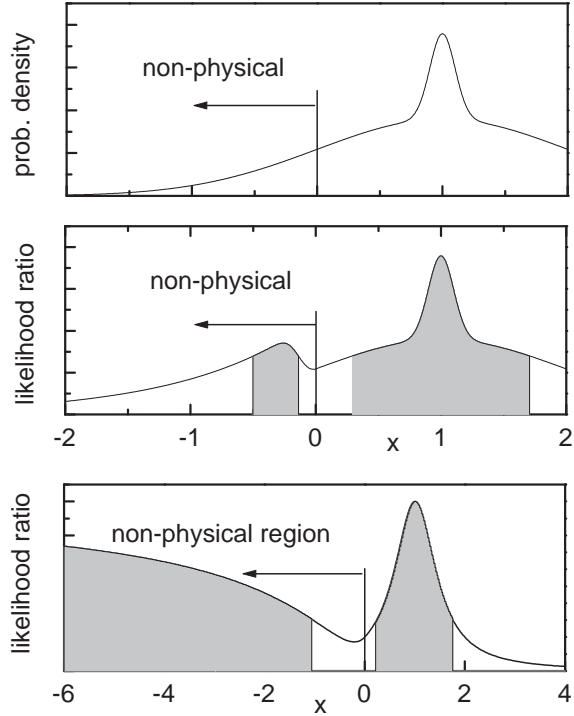}%
\caption{{\small Disconected probability intervals in the unified approach.
Gaussians (top) and Breit-Wigner (bottom).}}%
\end{center}
\end{figure}

\subsection{Example 1b: Superposition of Gaussians in the unified approach}

The prescription for the construction of the probability intervals according
to the likelihood ratio ordering leads to disconnected interval regions when
the pdf has tails and cannot produce confidence intervals. This is shown in
Fig. 4 top.

\subsection{Example 1c: Breit-Wigner distribution}

The same difficulty arises for the Breit-Wigner distribution (see Fig. 4 bottom).

The problem is absent if the pdf $f$ fulfills the condition $d^{2}\ln
f/dx^{2}\geq0$. This condition restricts the application of the unified
approach to pdfs similar to Gaussians.

\subsection{Example 2: Gaussian in two dimension and physical boundary}

Let us assume that we have a Gaussian resolution in $x,y$ and a physical
boundary in $y$ (Fig. 5). The probability contours are deformed in the
unified approach as indicated in the sketch. As a consequence the error in $x
$ shrinks due to a boundary in $y$ even though the two parameters are
independent. One has to be careful in the interpretation of two-dimensional
confidence limits as they occur for example in neutrino oscillation experiments.%

\begin{figure}
\begin{center}
\includegraphics[
width=8.5cm
]%
{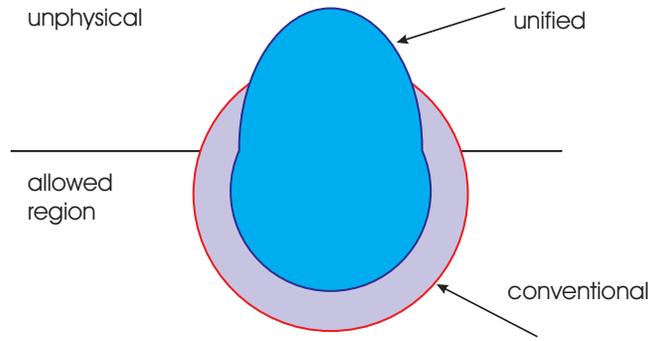}%
\caption{{\small Probabilty contours (schematic) for a two-dimensional
Gaussian near a boundary in the unified approach.}}%
\end{center}
\end{figure}

\subsection{Example 3: Slope of a linear distribution}

This is a frequent distribution in particle physics. A linear distribution is
always restricted in the sample and the parameter space to avoid negative
probabilities. We choose
\[
f(x|\theta)=\frac{1}{2}(1+\theta x);\hspace{1cm}-1\leq\theta,x\leq1
\]

as is realized in many asymmetry distributions. For a sample of 100 events
following the distribution of Equ. 2.4, a likelihood analysis gives a best
value for the slope parameter of $\hat{\theta}=0.92$ (see Figure 6). There
is no simple statistic allowing to compute central classical $62.8\%$
confidence limits because the parameter is undefined outside the interval
[1,1]. Contrary to the conventional classical approach, the unified approach
is able to handle the problem by working in the full sample space (hundred
dimensional in our case) This requires a considerable computing
effort\footnote{In my presentation at the meeting I had not realized this
solution in the unified approach. I thank Fred James and Gary Feldman for
explaining it to me.}.

Likelihood limits are possible - the upper limit would coincide with the
boundary - but not well suited to measure the precision.%

\begin{figure}
\begin{center}
\includegraphics[width=12.5cm
]%
{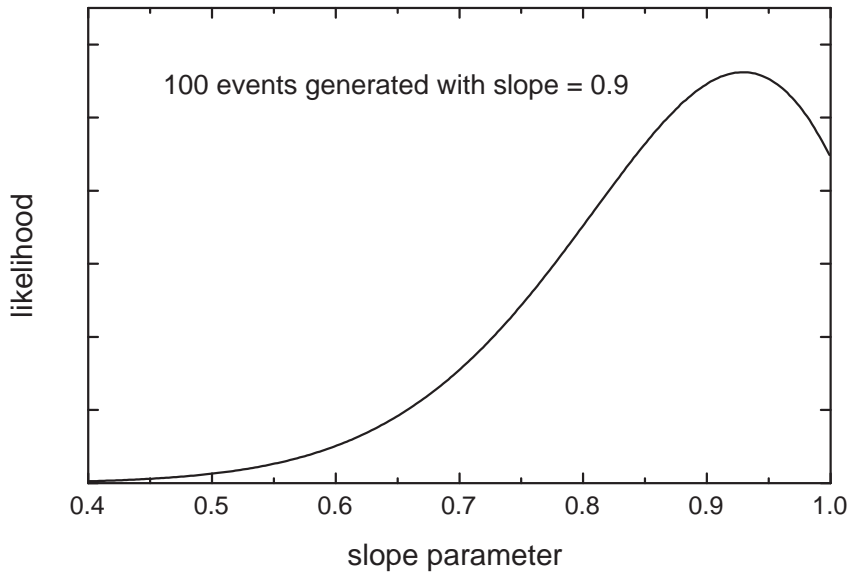}%
\caption{{\small Likelihood for a slope parameter.}}%
\end{center}
\end{figure}

\subsection{Example 4: Digital measurements}

A particle track is passing at the unknown position $\mu$ through a
proportional wire chamber. The measured coordinate $x$ is set equal to the
wire location\ $x_{w}$. The probability density for a measurement $x$%
\[
f(x,\mu)=\delta(x-x_{w})
\]
is independent of the true location $\mu$. Thus it is impossible to define a
sensible classical confidence or likelihood interval, except a trivial one
with full overcoverage. This difficulty is common to all digital measurements
because they violate condition 1 of section 2.1. Thus a large class of
measurements is not handled in classical statistics. A Bayesian treatment with
uniform prior is the common solution. It provides the r.m.s. error
$pitch/\sqrt{12}$.

\subsection{Example 5: Gaussian with two physical boundaries}%

\begin{figure}
[ptb]
\begin{center}
\includegraphics[
natheight=7.539400in,
natwidth=11.169900in,
height=2.2165in,
width=3.275in
]%
{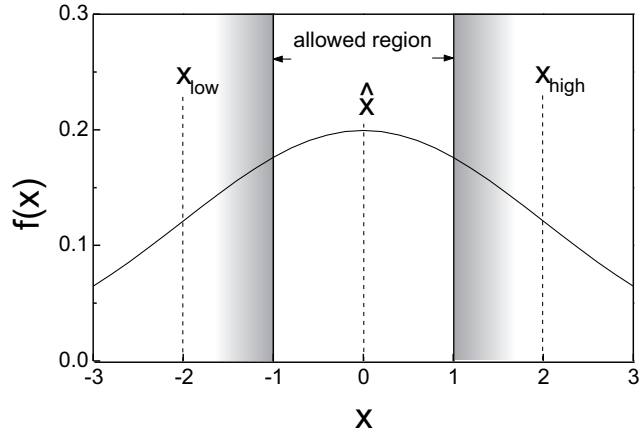}%
\caption{{\small Two-sided physical boundary. Classical error bounds cover the
full physical region.}}%
\end{center}
\end{figure}

A particle passes through a small scintillator and another position sensitive
detector with Gaussian resolution. Both boundaries of the classical error
interval are in the region forbidden by the scintillator signal. (see Fig.
7) The classical error is twice as large as the r.m.s. width. It is
meaningless. The unified classical and the likelihood limits contain the full
physical region and thus are useless. Again only the Bayesian method gives
reasonable results.

\subsection{Example 6: Gaussian with variable width}

A theory, depending on the unknown parameter $\theta$ predicts the Gaussian
probability density
\[
f(t)=\frac{25}{\sqrt{2\pi}\theta{}^{2}}\exp\left(  -\frac{625(t-\theta)^{2}%
}{2\theta^{4}}\right)
\]
for the time $t$ of an earthquake. The classical confidence interval for a
measurement at $t=10$ h is $7.66<t_{3}<\infty$. It is shown together with the
likelihood function in Fig. 8. When we look at the two distinct parameter
values, predicting the time of an earthquake%
\begin{figure}
\begin{center}
\includegraphics[width=14.5cm
]%
{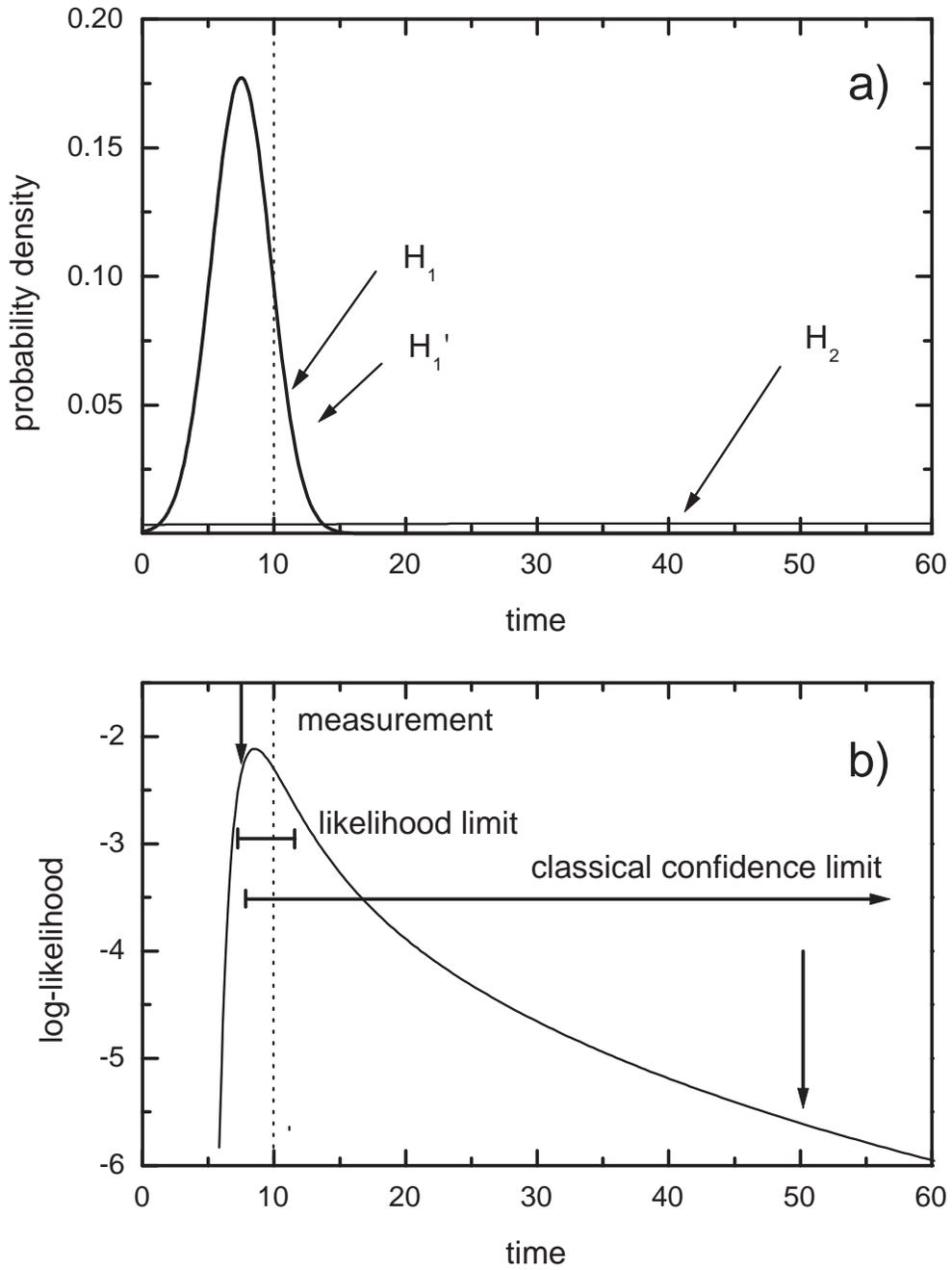}%
\caption{{\small Predictions from two discrete hypothesis H}$_{1}${\small ,
H}$_{2}${\small \ and measurement (a) and log-likelihood for parametrization
of the two hypothesis (b). The likelihood ratio strongly favors H}$_{1}%
${\small which is excluded by the classical confidence limits.}}%
\end{center}
\end{figure}
\begin{center}
\begin{tabular}{ll}
$H_1$: & $t_1=(7.50\pm2.25)$ h\\

$H_2$: & $t_2=(50\pm100)$ h
\end{tabular}
\end{center}
we realize that the first is excluded by the classical bounds, the second by
the likelihood limits. The Fig. 8b shows the two probability densities
together with the measurement. Clearly, we would rather accept $H_{1}$. This
choice is also supported by the likelihood ratio which is in favor of H$_{1}$
by a factor 26. Thus the likelihood limits are intuitively more acceptable
than the classical ones.

The preceding example shows that the concept of classical confidence limits
for continuous parameters is not compatible with methods based on the
likelihood values. We may construct a transition from the discrete case to the
continuous one by adding more and more hypothesis but a transition from
likelihood based methods to CCL is impossible. The two classical approaches
CCL and Neyman-Pearson test lack a common bases.

\subsection{Example 6b: Number of neutrinos}

This example was presented by Cousins \cite{cous95}: MarkII had measured the
number of neutrinos to be $2.8\pm0.6$ and deduced a 95\% confidence upper
limit of 3.9 excluding 4 neutrino generations. The likelihood ratio of 7.0
produces a much weaker exclusion of the discrete hypothesis.

\subsection{Example 7: Stopping rule}

A rate measurement may be stopped for reasons like: i) There are enough
events. ii) For a long time no event has been observed. iii) A ``golden''
event was recorded.%

\begin{figure}
\begin{center}
\includegraphics[width=8.5cm
]%
{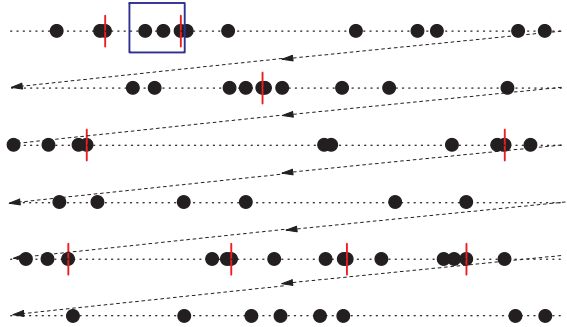}%
\caption{{\small A sequential stopping rule does not introduce a bias.}}%
\end{center}
\end{figure}

These actions do not introduce a bias as has been first realized by Barnard
and co-workers \cite{barn62}. The reason is that the likelihood function is
independent of the stopping rule. This may be visualized by an infinitely long
measurement which is cut in pieces each corresponding to a experiment stopped
by the same rule. The individual experiment cannot be biased since the full
chain is unbiased. This is illustrated in Fig. 9 where the experiments are
stopped whenever 3 events are recorded in a short time interval.%

\begin{figure}
\begin{center}
\includegraphics[width=12.5cm
]%
{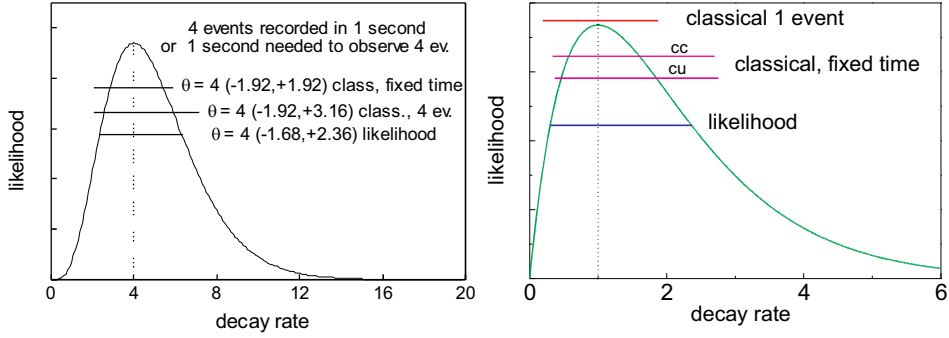}%
\caption{{\small Stopping after a fixed time or when a fixed number of events
has been observed (same likelihood) gives different results in classical
statistics.}}%
\end{center}
\end{figure}

The Figure 10 shows the likelihood function for an experiment where 4 events
are observed in a time interval of one second. The classical results depend on
the stopping condition: a) the time interval had been fixed, b) the experiment
was stopped after the forth event. The likelihood principle states that the
two data sets are equivalent. Thus the classical limits are inconsistent.

The differences become even larger when we take the example of 1 event
recorded in 1 second (see Fig. 10 right). The likelihood functions given by
the lifetime distribution and the Poisson distribution, respectively are
proportional to each other
\begin{eqnarray*}
f(t|\lambda)  &  =\lambda e^{-\lambda t}\\
P(1|\lambda)  &  =\frac{e^{-\lambda}\lambda^{1}}{1!}%
\end{eqnarray*}

\subsection{Example 8: Poisson signal with background}

In a garden there are apple and pear trees. Usually during night some pears
fall from the trees. One morning looking from his window, the proprietor who
is interested in apples find that no fruit is lying in the grass. Since it is
still quite dark he is unable to distinguish apples from pears. He concludes
that the average rate of falling apples per night is less the 2.3 with 90\%
confidence level. His wife who is a classical statistician tells him that his
rate limit is too high because he has forgotten to subtract the expected pears
background. He argues, ``there are no pears'', but she insists and explains
him that if he ignores the pears that could have been there but weren't, he
would violate the coverage requirement. In the meantime it has become bright
outside and pears and apples - which both are not there - are now
distinguishable. Even though the evidence has not changed, the classical limit has.

The 90\% confidence limits for zero events observed and background expectation
$b=0$ is $\mu=2.3$. For $b=2$ it is $\mu^{\prime}=0.3$ much lower. \emph{CCL
are different for two experiments with exactly the same experimental evidence
relative to the signal (no signal event seen)}. This situation is absolutely
intolerable. Feldman and Cousins consider this kind of objections as ``based
on a misplaced Bayesian interpretation of classical intervals'' \cite{feld98}.
It is hard to detect a Bayesian origin in a generally accepted principle in
science, namely, two measurements containing the same information should give
identical results. The critics here is not that CCLs are inherently wrong but
that their application\textbf{\ }to the computation of upper limits when
background is expected does not make sense, i.e. these limits do not measure
the precision of the experiment.

The effect is less dramatic but also present in the unified approach: An
experiment finding no event n=0 with background expectation b=3 produces a
90\% confidence limit 1.08 for the signal (see Table 2.1). Then the flux is
doubled and the background is eliminated. The limit becomes 2.44/2=1.22, worse
than before. This problem is absent in the versions proposed by Roe and
Woodroofe \cite{roe99} and also in that of Punzi \cite{punz99}. These methods
are however restricted to the Poisson case.%

\begin{table}[tbp] \centering
\begin{tabular}
[c]{|l|l|l|l|l|l|}\hline
& n=0, b=0 & n=0, b=1 & n=0, b=2 & n=0, b=3 & n=2, b=2\\\hline
standard classical & 2.30 & 1.30 & 0.30 & -0.70 & 3.32\\\hline
unified classical & 2.44 & 1.61 & 1.26 & 1.08 & 3.91\\\hline
uniform Bayesian & 2.30 & 2.30 & 2.30 & 2.30 & 3.88\\\hline
\end{tabular}
\caption{Poisson limits in classical and Bayesian approaches\label{key}}%
\end{table}%

To avoid the unacceptable situation, I have proposed a modified frequentist
approach to the calculation of the Poissonian limits including the information
of the limited number of background events \cite{zech89}. There the confidence
level is normalized to the probability to observe $0\leq n_{b}\leq n$
background events as known from the measurement.
\[
1-\alpha=\frac{\sum_{i=0}^{n}P(i|\mu+b)}{\sum_{i=0}^{n}P(i|b)}%
\]
The resulting limits respect the likelihood principle (see below) and thus are
consistent. They coincide with those of the uniform Bayesian method and
provide a frequentist interpretation of the Bayesian limits. However, as has
been pointed out by Highland \cite{high97}, the limits do not have minimum
overcoverage as required by the strict application of the Neyman construction.
This is correct \cite{zech97} but in my paper no claim relative coverage had
been made. The method has been applied to the a Higgs search \cite{read97}.%

\begin{figure}
\begin{center}
\includegraphics[width=14.0cm
]%
{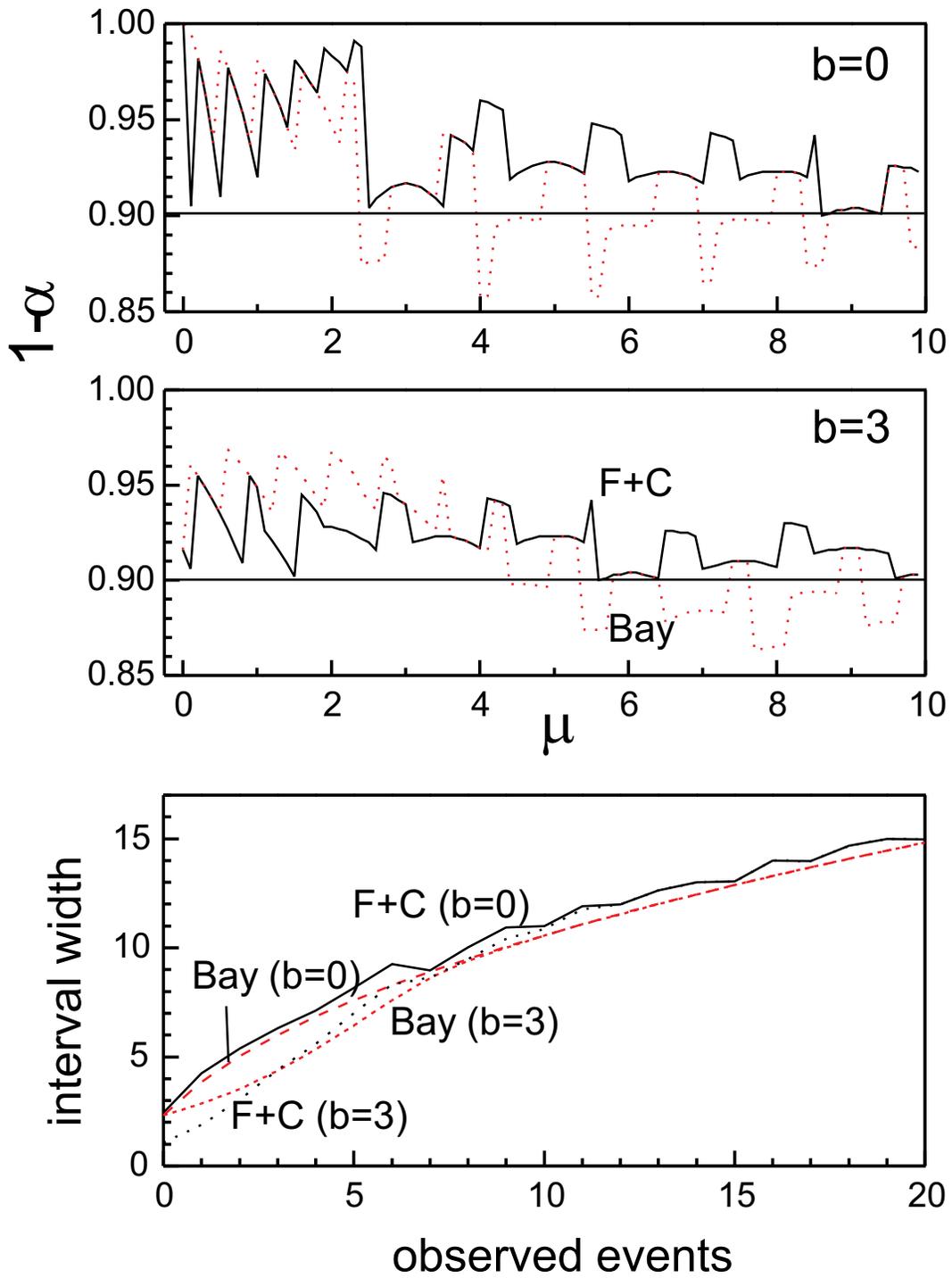}%
\caption{{\small Coverage in the unified classical and the Bayesian approach
(dotted) and interval lengths (bottom).}}%
\end{center}
\end{figure}

Often the background expectation is not known precisely since it is estimated
from side bands or from other measurements with limited statistics. So far,
there is no classical recipe which allows to incorporate an uncertainty of the
background estimate.

Likelihood limits also give a sensible description of the data. Whether
likelihood limits or Bayesian limits obtained from the integration are more
sensible depends on the shape of the likelihood function. Ideally both limits
should be given.

Fig. 11 compares the coverage of the unified classical and the Bayesian
limits. At small signals both overcover strongly. For large signals the
Bayesian method slightly undercovers and oscillates around the nominal value.

\subsection{Example 9: Combining lifetime measurements}

Two events are observed from an exponential decay with true mean life
$\tau_{0}=1/\gamma_{0}$. The maximum likelihood estimate is used either for
$\tau$ or $\gamma$. We assume that an infinite number of identical experiments
is performed and that the results are combined. In Table 2.2 we summarize the
results of different averaging procedures. There is no prescription for
averaging classical intervals. The unified methods have to explain how they
intend to combine their measurements. To compute the classical result given in
the table, the maximum likelihood estimate and central intervals were used.%

\begin{table}[tbp] \centering
\begin{tabular}
[c]{|lll|}\hline
\multicolumn{1}{|l|}{method} & $<\tau/\tau_{0}>$ & $<\gamma/\gamma_{0}%
>$\\\hline
\multicolumn{1}{|l|}{adding log likelihood functions} & $1$ & $1$\\
\multicolumn{1}{|l|}{classical, weight: $\sigma^{-2}$} & $0.0$ & $0.67$\\
\multicolumn{1}{|l|}{likelihood, weight: PDG} & $0.26$ & $0.80$\\
\multicolumn{1}{|l|}{Bayesian mean, uniform prior, weight: $\sigma^{-2}$} &
$\infty$ & $1$\\\hline
\end{tabular}
\caption{Average of an infinite number of equivalent
lifetime measurements using different weighting procedures}%
\end{table}

\begin{figure}
[ptb]
\begin{center}
\includegraphics[
trim=0.000000pt 23.814650pt 0.000000pt 0.000000pt,
natheight=226.375000pt,
natwidth=319.187500pt,
height=121.9375pt,
width=191.375pt
]%
{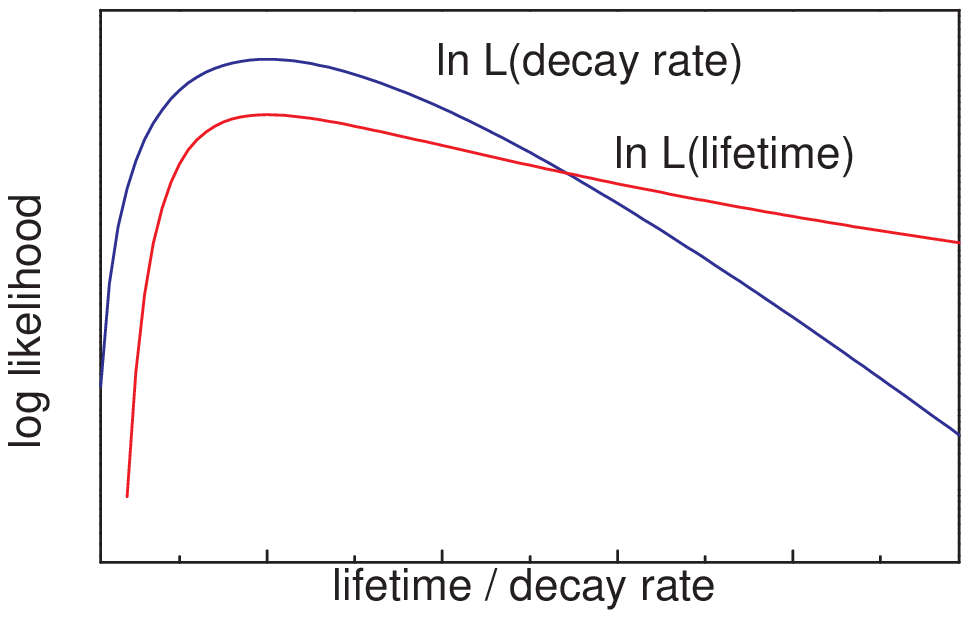}%
\caption{{\small Log-likelihood function of the mean life and the decay
rate.}}%
\end{center}
\end{figure}

In this special example a consistent result is obtained in the Bayesian method
with uniform prior for the decay constant. It shows also how critical the
choice of the parameter is in the Bayesian approach. It is also clear that an
educated choice is also important for the pragmatic procedures. It is obvious
that the decay constant is the better parameter (see also Fig. 12). Methods
approximating the likelihood function provide reasonable results unless the
likelihood function is very asymmetric. The weighting procedure of the PDG
applied to the likelihood errors gives reasonable results. As is well known,
adding the log-likelihood functions always produces a correct result.

\section{Conclusions}

\subsection{Conventional classical method}

The conventional classical schemes suffer from the following problems:

\begin{itemize}
\item  There are inconsistencies ( Poisson limits, stopping rule, discrete vs.
continuous parameters).

\item  There is a lack of precision (unphysical limits).

\item  They have a restricted range of application (problems with digital
measurements, discrete parameters).

\item  They are not invariant against sample variable transformations (except
central intervals in one dimension).

\item  They are subjective (coverage requires pre-experimental fixing of cuts
and decision to publish).

\item  There are unsolved problems. (It is not clear how to combine
measurements. The inclusion of background errors in Poisson processes is not possible.)

\item  There is no obvious treatment of nuisance parameters.

\item  Systematic errors cannot be included.
\end{itemize}

\subsection{Unified approach}

Compared to the conventional method there are improvements:

\begin{itemize}
\item  The inconsistencies in Poisson processes are weaker ( and absent in the
version of Roe and Woodroofe)

\item  Non-physical limits are avoided.

\item  It is invariant with respect to variable and parameter transformations.
\end{itemize}

However most problems remain (inconsistencies, lack of precision, background
uncertainty in Poisson limits), and:

\begin{itemize}
\item  It is restricted to specific pdfs (Gaussian like).

\item  It is complicated and requires considerable computing efforts.

\item  The combination of measurements is even more unclear.

\item  Artificial error correlations are introduced near boundaries.

\item  The proposed treatment\ \cite{cous00} of nuisance parameters (use best
estimate may lead to undercoverage.
\end{itemize}

\subsection{Likelihood limits}

Likelihood limits have\ attractive properties

\begin{itemize}
\item  They are consistent.

\item  They provide optimum precision.

\item  They are invariant against variable and parameter transformations.

\item  They provide a coherent transition to discrete hypothesis (likelihood ratio)

\item  Measurements can easily be combined
\end{itemize}

There are also restrictions in the application:

\begin{itemize}
\item  Digital measurements and uniform distributions cannot be handled.
\end{itemize}

\subsection{Bayesian limits}

The Bayesian philosophy is very general and flexible:

\begin{itemize}
\item  All problems can be treated. (Nuisance parameters, digital
measurements, unphysical boundaries etc.)
\end{itemize}

but:

\begin{itemize}
\item  They depend on the parameter choice.
\end{itemize}

\section{Proposed conventions}

The conventions proposed here represent by no means the only reasonable prescription.

Since the complete information is contained in the likelihood function,
classical approaches are not considered. (They cannot be computed from the
likelihood function alone.) An even stronger reason for there exclusion are
the obvious inconsistencies of this method.

The main objection against Bayesian methods is their dependence on the
selected parameter. I find it rather natural to choose a sensible parameter
space. For some application like pattern recognition - which, by the way,
cannot be done with classical statistics - it is absolutely necessary.

The proposed conventions are:

\begin{enumerate}
\item  Whenever possible the full likelihood function should be published. It
contains the experimental information and permits to combine the results of
different experiments in an optimum way. This is especially important when the
likelihood is strongly non-Gaussian (strongly asymmetric, cut by external
bounds, has several maxima etc.).

\item  Data are combined by adding the log-likelihoods. When not known,
parametrizations are used to approximate it.

\item  If the likelihood is smooth and has a single maximum the likelihood
limits should be given to define the error interval. These limits are
invariant under parameter transformation. For the measurement of the parameter
the value maximizing the likelihood function is chosen. No correction for
biased likelihood estimators is applied. The errors usually are asymmetric.
These limits can also be interpreted as Bayesian one standard deviation errors
for the specific choice of the parameter variable where the likelihood of the
parameter has a Gaussian shape.

\item  Nuisance parameters are eliminated by integrating them out using an
uniform prior. A correlation coefficient should be computed.

\item  For digital measurements the Bayesian mean and r.m.s. should be used.

\item  In cases where the likelihood function is restricted by physical or
mathematical bounds and where there are no good reasons to reject an uniform
prior the measurement and its errors defined as the mean and r.m.s. should be
computed in the Bayesian way.

\item  Upper and lower limits are computed from the tails of the Bayesian
probability distributions. (In some cases likelihood limits may be more
informative. \cite{dago00})

\item  Non-uniform prior densities should not be used.

\item  It is the scientist's choice whether to present an error interval or an
upper limit.

\item  In any case the applied procedure has to be documented.
\end{enumerate}

These recipes correspond more or less to our every day practice. An exception
are Poisson limits where for strange reasons the coverage principle - though
only approximately realized - has gained preference in neutrino experiments.

\textbf{Acknowledgement}

I would like to thank Fred James, Louis Lyons and Yves Perrin for having
organized this interesting workshop which - for the first time - offered the
possibility to high energy physicists to expose and discuss their problems and
solutions to statistical problems.


\begin{thebibliography}{99}
\bibitem{nars99}I. Narsky, Estimation of Upper Limits Using a Poisson
Statistic, hep-ex/9904025 (1999)

\bibitem{edwa92}A. W. F. Edwards, Likelihood, The John Hopkins University
Press, Baltimore (1992)

\bibitem{zech98}G. Zech, Objections to the unified approach to the computation
of classical confidence limits, physics/9809035

\bibitem{feld98}G. J. Feldman, R. D. Cousins, Unified approach to the
classical statistical analysis of small signals. Phys. Rev. D 57 (1998) 1873.

\bibitem{basu88}D. Basu, Statistical Information and Likelihood, Lecture Notes
in Statistics, ed. J.K. Ghosh, Springer-Verlag NY (1988)

\bibitem{berg84}J. O. Berger and R. L. Wolpert, The likelihood Principle,
Lecture Notes of Inst. of Math. Stat., Hayward, Ca, (ed. S. S. Gupta) (1984)

\bibitem{cous95}R. D. Cousins, Why Isn't Every Physicist a Bayesian? Am J.
Phys. 63 (1995) 398

\bibitem{barn62}G. A. Barnard, G. M. Jenkins and C. B. Winstein, Likelihood
inference and time series, J. Roy. Statist. Soc. A 125 (1962)

\bibitem{roe99}B. P. Roe and M. B. Woodroofe, Improved probability method for
estimating signal in the presence of background, Phys. Rev. D 60, 053009 (1999)

\bibitem{punz99}G. Punzi, A stronger classical definition of confidence
limits, hep-ex/9912048 (1999)

\bibitem{zech89}G. Zech, Upper limits in experiments with background or
measurement errors, Nucl. Instr. and Meth. A277 (1989) 608-610

\bibitem{high97}V. L. Highland, Comments on ``Upper limits in experiments with
background or measurement errors'', Nucl. Instr. and Meth. A 398 (1997) 429

\bibitem{zech97}G. Zech, Reply to 'Comments on ``Upper limits in experiments
with background or measurement errors'' ', Nucl. Instr. and Meth. A398 (1997) 431-433

\bibitem{read97}A. L. Read, Optimal statistical analysis of search results
based on the likelihood ratio and its application to the search for the MSM
Higgs boson at $\sqrt{s}$ = 161 and 172 GeV, DELPHI 97-158 PHYS 737 (1997)

\bibitem{dago00}G. D'Agostini, Contribution to this workshop

\bibitem{cous00}R. D. Cousins, Contribution to this workshop
\end{thebibliography}
\end{document}